\documentclass{PoS}

\usepackage{latexsym}
\usepackage{subfigure}
\usepackage{multicol}
\usepackage{multirow}
\usepackage{mathrsfs}
\usepackage{verbatim}
\usepackage{amsmath,amsthm, amssymb}

\title{Numerical investigations of Supersymmetric Yang-Mills Quantum Mechanics with 4 supercharges} 
\ShortTitle{Numerical investigations of SYMQM with 4 supercharges}

\author{
Zbigniew Ambrozi\'nski $^a$, \speaker{Piotr Korcyl } $^{ab}$ \hfill{\footnotesize{\it DESY 14-223}}\\
\llap{$^a$}
M. Smoluchowski Institute of Physics, Jagellonian University\\
ul. \L ojasiewicza 11, 30-348 Krak\'ow, Poland\\
\llap{$^b$} John von Neumann Institute for Computing (NIC), DESY\\ 
Platanenallee 6, D-15738 Zeuthen, Germany\\
E-mail: \email{piotr.korcyl@desy.de}}

\abstract{
We report on our non-perturbative investigations of supersymmetric Yang-Mills quantum mechanics with 4 supercharges.
We employ two independent numerical methods. First of them is the cut Fock space method whose numerical implementation 
was recently generalized to include the SU(N) gauge group. It allowed us to calculate for the first time the spectrum of the 
model with SU(3) symmetry in all fermionic sectors. Independently, we implemented the Rational Hybrid Monte Carlo algorithm 
and reproduced the accessible part of the low-energy spectrum of the model with SU(2) gauge symmetry. 
We argue that by simulating at imaginary chemical potential one can get access to observables defined in sectors of Hilbert 
space with a given quark number. 
}

\FullConference{The 32nd International Symposium on Lattice Field Theory\\
                 23-28 June, 2014\\
                 Columbia University New York, NY}

\newcommand{\tr}{\mathrm{tr}\,}

\begin{document}

According to one of the generalizations of the AdS/CFT conjecture one can relate supersymmetric
Yang-Mills quantum mechanics with 16 supercharges to a certain string theory describing
black holes \cite{banks}. This result motivated several interesting non-perturbative numerical 
simulations which studied thermodynamical properties of this system \cite{many}.
Assuming that the mentioned conjecture is correct, the microscopic details of the spectrum of this model, for example
the wavefunction of the ground state, might turn to be useful on our way towards the understanding of physics 
of quantum black holes. In this Letter we report on two numerical approaches which can
provide this type of information. Both of them are applied to supersymmetric Yang-Mills quantum mechanics with 4 supercharges,
which may be thought of as a simplified version of the model with 16 supercharges and which we use 
as a test ground for our methods. We briefly introduce this model in Section \ref{sec. intro}.
First of the two approaches is based on the Hamiltonian formulation of supersymmetric 
Yang-Mills quantum mechanics. The recursive algorithm, which is necessary to reduce the computation time to 
menageable timescales \cite{campostrini},
was recently generalized to any SU(N) gauge group \cite{zbyszek} \cite{phd}. It allowed us to compute for the first time 
the eigenenergies and eigenstates of the model with SU(3) symmetry and to disentangle the discrete spectrum from the continuous one. 
Our second approach uses traditional Monte Carlo simulations. We argue that by using lattice simulations one can extract 
few lowest eigenenergies with a precision comparable to the Hamiltonian approach and moreover that by simulating at imaginary 
chemical potential one has access to sectors of the Hilbert space with a given fermionic occupation number. 
In the following we describe our progress along these lines: in Section \ref{sec. fock} we report on our new results 
obtained with the Hamiltonian approach and in Section \ref{sec. lattice} we describe the details of our Monte Carlo simulations. 
Eventually, Section \ref{sec. conclusions} contains conclusions. 



\section{Supersymmetric Yang-Mills Quantum Mechanics with 4 supercharges}
\label{sec. intro}

The euclidean action of supersymmetric Yang-Mills quantum mechanics with 4 supercharges
\begin{equation}
S = \tr \int_R d\tau \Big\{ \frac{1}{2}(D_{\tau} X_i)^2 - \frac{1}{4}\Big[ X_i, X_j \Big]^2 + \Psi^{\dagger} D_{\tau} \Psi - \Psi^{\dagger} \sigma^i \Big[ X_i, \Psi \Big] \Big\}
\label{eq. action}
\end{equation}
where $i=1,2,3$ and $\Psi$ are complex, two-component Grassmann variables, can be obtained by dimensional 
reduction of $\mathcal{N}=1$, $D=4$ supersymmetric Yang-Mills quantum field theory. The coupling constant 
$g$ can be factored out and we set $g=1$ for simplicity. The scalar fields transform in the
adjoint representation of the gauge group which in this case is reduced to a time-dependent symmetry. 
$D_{\tau}$ is the covariant derivative given by
\begin{equation}
D_{\tau} X_i = \partial_{\tau} X_i + i[A_0(\tau), X_i].
\end{equation}
The Hamiltonian 
\begin{equation}
H = \tr \Big\{ P_i^2 - \frac{1}{2} \Big[ X_i, X_j \Big]^2 + \Psi^{\dagger} \sigma^i \Big[ X_i, \Psi \Big] \Big\}
\end{equation}
commutes with the gauge invariant quark number operator,
\begin{equation}
Q = \tr \Psi^{\dagger}_{\alpha} \Psi_{\alpha} \textrm{ with integer eigenvalues: } q = 0, \dots ,6. 
\end{equation}
We are interested in the spectral properties of this model. 

\section{Cut Fock space approach: SU(2) and SU(3) models}
\label{sec. fock}

The quark number $n_F$ being conserved, the Hilbert space decomposes into a direct sum of Hilbert spaces defined in given fermionic sectors. 
A recursive algorithm can be set up to construct a Fock basis using gauge invariant creation operators in each of them. 
All redundant states are eliminated through an orthogonalization step. The Hamiltonian and the angular momentum operator matrices can 
then be calculated and simultaneously diagonalized. As we include basis states with increasing number of bosonic quanta convergence of the 
eigenenergies can be seen. The model with SU(2) symmetry was investigated with this method several years ago \cite{campostrini}. 
Recently, the recursive
algorithm was generalized to include higher SU(N) groups and for the first time the spectrum of the SU(3) model was computed 
\cite{zbyszek} \cite{phd}. 

\begin{figure}
\begin{center}
\includegraphics[width=0.55\textwidth]{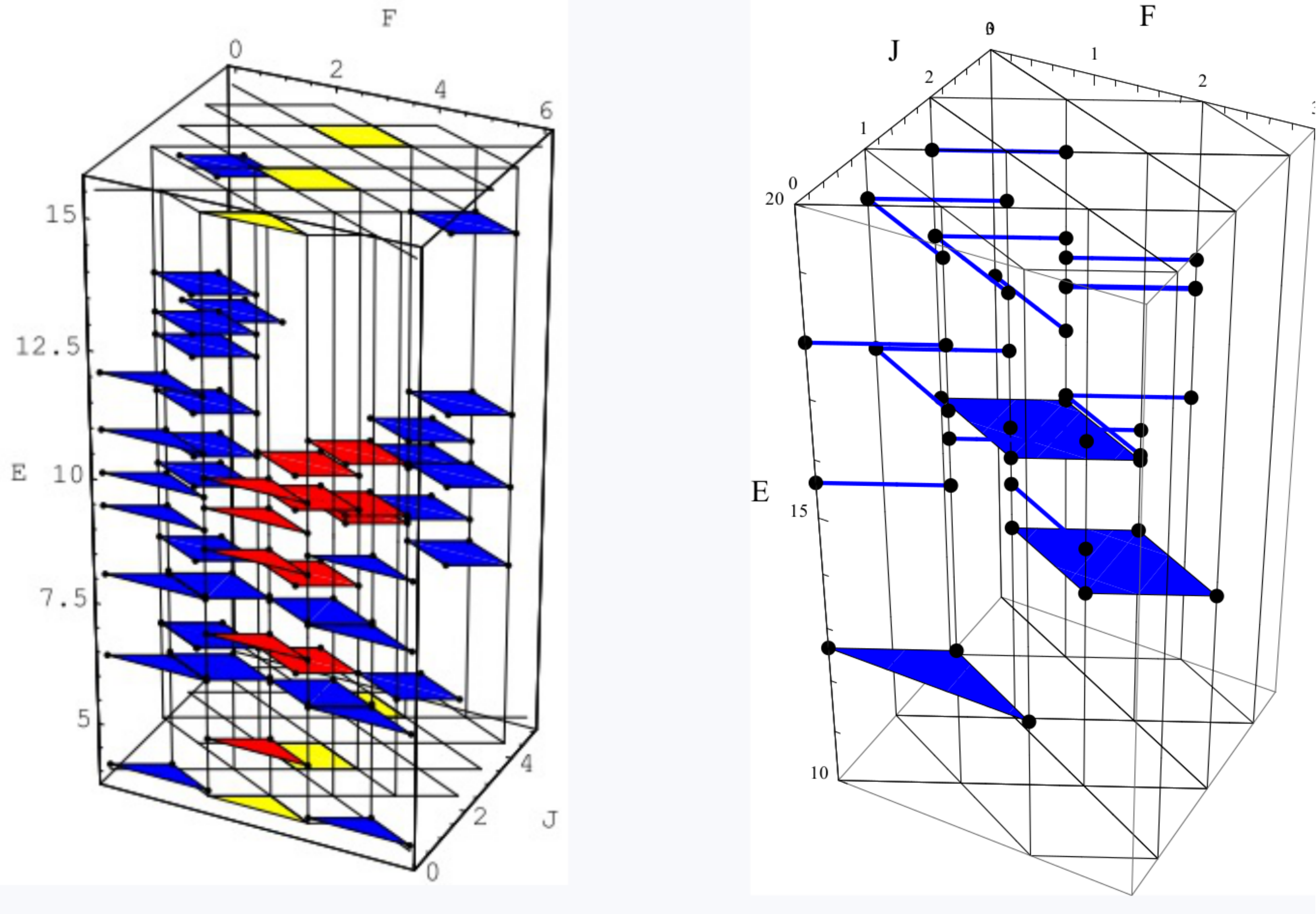}
\caption{Supersymmetric structure of the spectrum for the SU(2) \cite{campostrini} and SU(3) \cite{zbyszek} \cite{phd} models \label{fig. spectra}}
\end{center}
\end{figure}

On figure \ref{fig. spectra} we show a sample of results obtained from this approach, i.e. the 
structure of the spectrum as a function of fermionic occupation number and angular momentum for the SU(2) model (left panel) and for the
SU(3) model. Blue diamonds denote fully identified supermultiplets, namely 4 eigenstates belonging to the discrete spectrum and lying 
in different sectors but having degenerate energies. The red diamonds are used to picture the identified supermultiplets of bound states immersed in the 
continuous spectrum. For the SU(3) model we expect that the sectors with continuous spectrum have at least $n_F=5$. These sectors are not
shown since within our actual precision we couldn't fully identify any supermultiplets.
%
%


\section{Lattice approach: SU(2) model}
\label{sec. lattice}

We discretize the action Eq.\eqref{eq. action} on a circle of unit circumference with $M$ sites in the following way
\begin{multline}
S = \frac{1}{\beta^3} \sum_{a=0}^{M-1} \tr \Big\{ \frac{1}{2\epsilon} 
\Big( X_{i,a} - U_a X_{i,a-1} \Big)^2 - \frac{\epsilon}{4} \Big[ X_{i,a}, X_{j,a}\Big]^2 + \\
+ \Psi^{\dagger}_{\alpha, a} \Big( \Psi_{\alpha, a} - U_a \Psi_{\alpha, a-1} \Big)
-\epsilon \Psi^{\dagger}_{\alpha, a} \sigma^{i,\alpha \beta} \Big[ \Psi_{\beta, a}, X_{i,a}\Big]\Big\}
\end{multline}
We have defined dimensionless fields by rescaling $X_{i,a} \rightarrow R^{-1} X_{i,a}$ and 
$\Psi_{\beta,a} \rightarrow R^{-\frac{3}{2}} \Psi_{\beta,a}$. The overall $R^{-3}$ factor
is written as $\beta^{-3}$ following the conventions introduced in Ref.\cite{catterall}. The 
dimensionless lattice spacing $\epsilon$, satisfying $M \epsilon = 1$, can be factored out 
from the bosonic part of the action by the rescaling $X_i \rightarrow \frac{1}{\epsilon} X_i$. 
The links $U_a$ are in the adjoint representation. We imposed antiperiodic boundary conditions 
for the fermionic fields and periodic boundary conditions for the scalar fields. The final form 
of the action which we implemented reads
\begin{multline}
S = \frac{1}{\beta^3} \sum_{a=0}^{M-1} \tr \Big\{ M^3 \Big( \frac{1}{2} 
\big( X_{i,a} - U_a X_{i,a-1} \big)^2 - \frac{1}{4} \big[ X_{i,a}, X_{j,a}\big]^2 \Big) + \\
+ \Big( \Psi^{\dagger}_{\alpha, a} \big( \Psi_{\alpha, a} - U_a \Psi_{\alpha, a-1} \big)
- \Psi^{\dagger}_{\alpha, a} \sigma^{i,\alpha \beta} \big[ \Psi_{\beta, a}, X_{i,a}\big]\Big\}
\end{multline}
We used the Metropolis and HMC algorithms to simulate the bosonic part of the action and the 
RHMC with reweighting algorithm to simulate the full theory. The continuum limit can be simply 
taken by $M\rightarrow \infty$ while keeping $\beta$ fixed. We use the fact that 
$\langle S_{\textrm{fermionic}} \rangle = 2M(N^2-1)$ to check that our runs are thermalized.

Among the physical observables that we measure (on a lattice with $M$ sites) are:
\begin{itemize}
\item dimensionless bosonic ground state energy $R E_0(M)$ defined as
\begin{equation}
R E_0 (M) = \frac{3}{4} \frac{M^2}{\beta^3} \sum_a \langle \Big[ X_{i,a}, X_{j,a}\Big]^2 \rangle
\end{equation}
where we used the quantum virial theorem to trade the mean value of the kinetic energy operator for 
the mean value of the potential energy operator,
\item the correlator $C_1(n)$
\begin{equation}
C_1(n) = \langle X_{i,a} \prod_{k=1}^n U_{a+k} X_{i,a+n} \rangle \sim e^{- R E_1(M) n},
\end{equation}
from which we estimate the energy of the first excited state $R E_1(M)$. Since $X_{i,a}$ can excite only one state, 
the correlator decays with a single exponential $e^{- R E_1(M) n}$; it is
not contaminated with higher excited states at small distances,
\item and the Polyakov loop defined as
\begin{equation}
P(M) = \langle |\tr e^{i\int A d\tau}| \rangle = \langle | \tr \prod_{a=1}^M U_a | \rangle.
\end{equation}
\end{itemize}

\subsection{Bosonic sector}

\begin{figure}
\begin{center}
\includegraphics[width=0.485\textwidth]{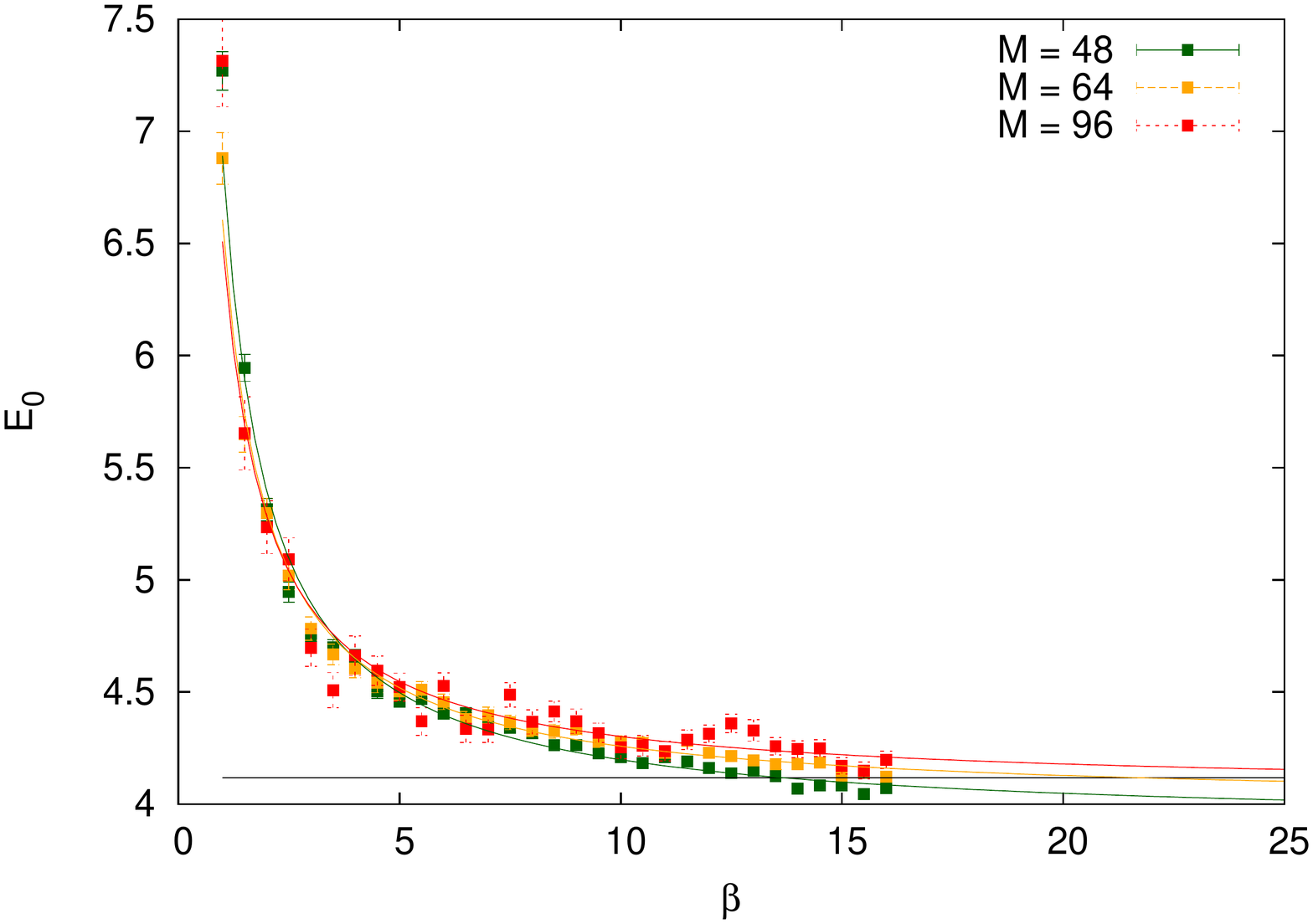}
\includegraphics[width=0.485\textwidth]{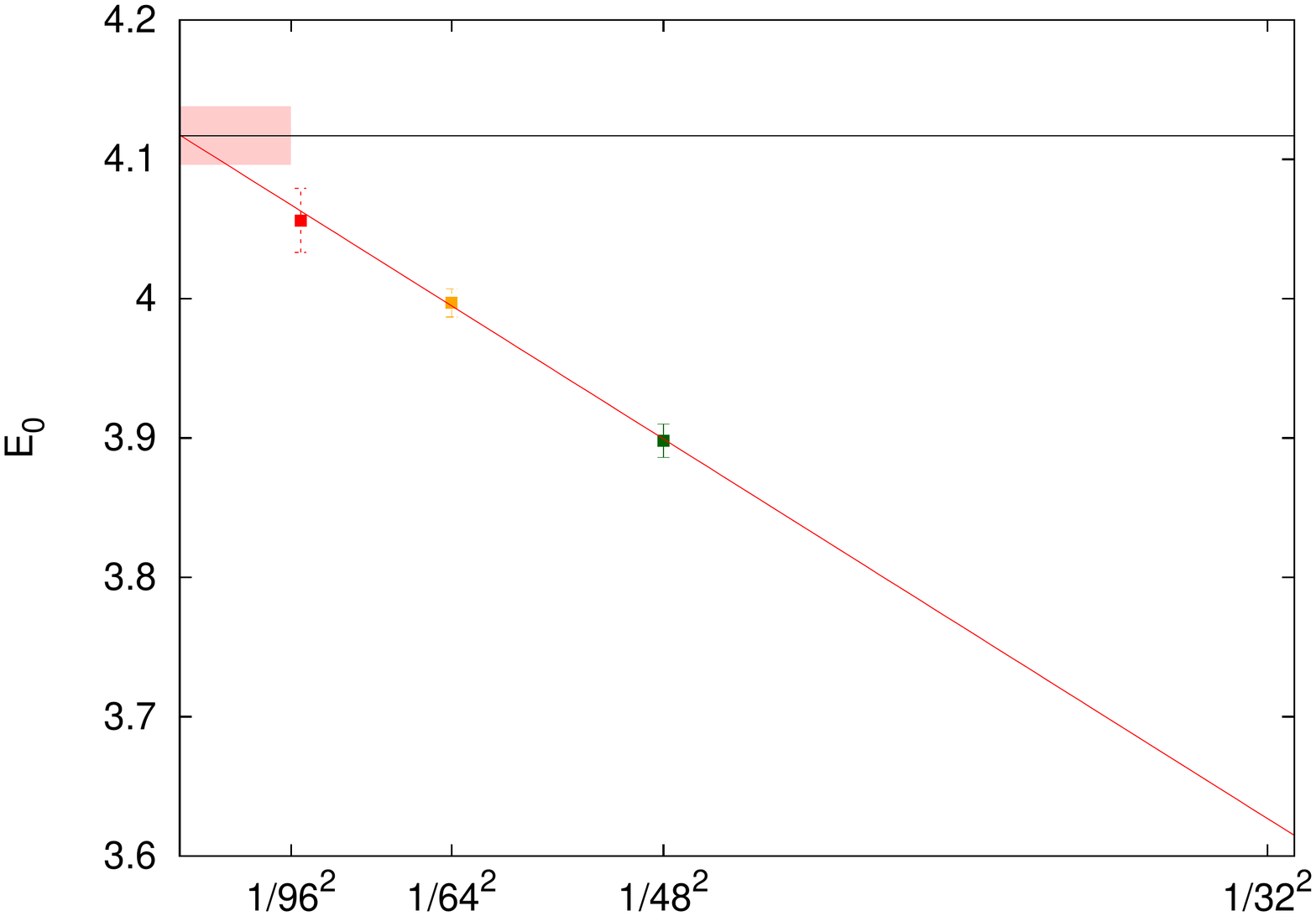}
\caption{Extraction of the ground energy: $E_0 = 4.117(21)$: the infinite volume (large $\beta$) 
extrapolation is shown on the left panel and the continuum extrapolation is shown on the right panel. 
The black solid line corresponds to the result from the Hamiltonian approach, whereas the pink area 
corresponds to the uncertainty of the lattice result.
 \label{fig. e0}}
\end{center}
\end{figure}

\begin{figure}
\begin{center}
\includegraphics[width=0.485\textwidth]{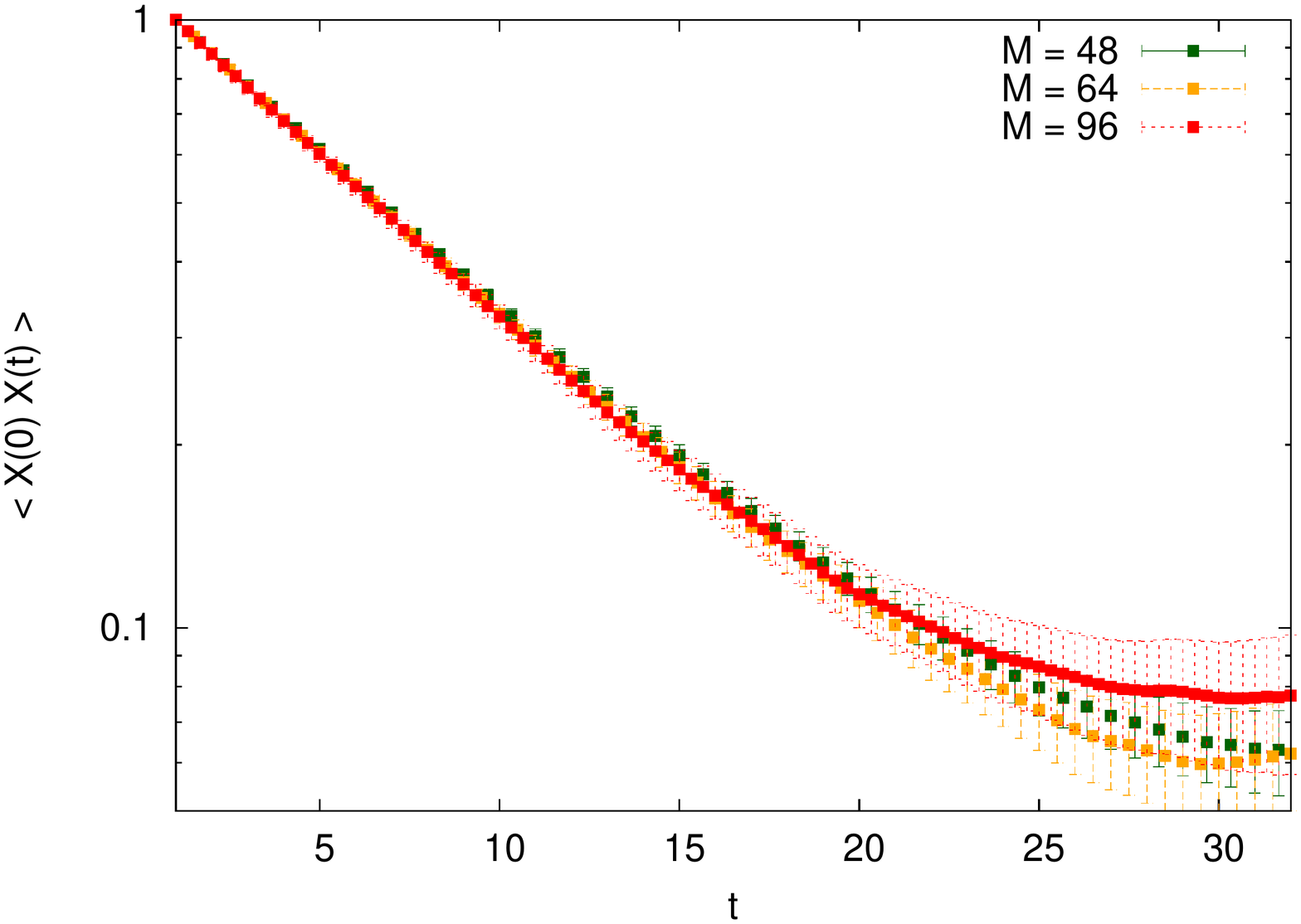}
\includegraphics[width=0.485\textwidth]{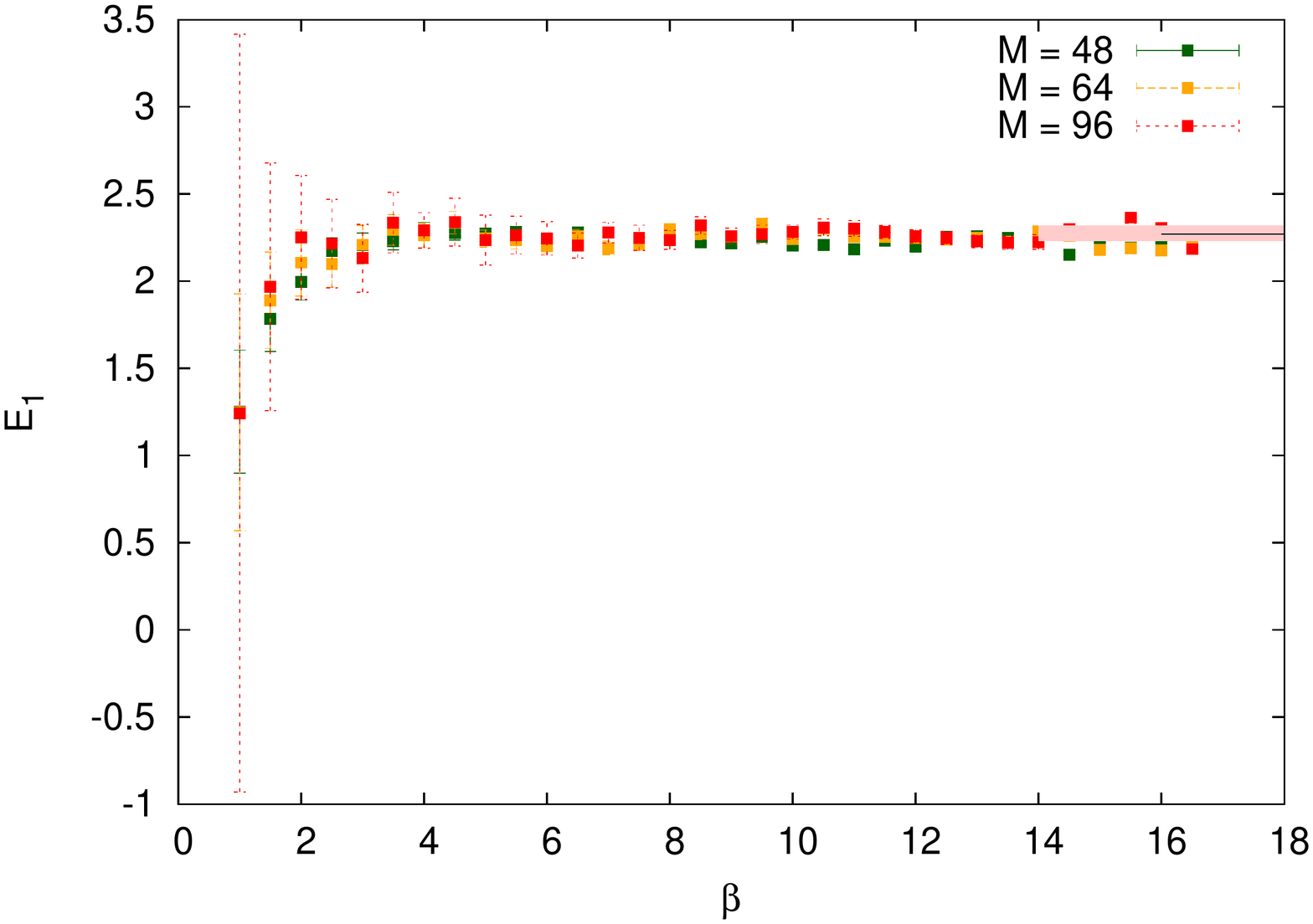}
\caption{Extraction of the first excited state energy: $E_1 = 6.397(71)$:
the left panel shows the correlator $C_1(n)$ for different lattice spacings whereas the right panel shows the infinite volume
extrapolation. The black solid line corresponds to the result from the Hamiltonian approach, whereas the pink area 
corresponds to the uncertainty of the lattice result.
 \label{fig. e1}}
\end{center}
\end{figure}

In this section we present our lattice simulations of the bosonic sector of the SU(2) model and compare
them with the results obtained with the Hamiltonian approach \cite{campostrini}. Figure \ref{fig. e0} shows
the results for the bosonic ground energy. On the left panel we demonstrate the infinite volume/zero temperature 
extrapolation which is needed to match to the energy obtained from the Hamiltonian approach. We use the
extrapolation ansatz for the dimensionfull energy $E_0$
\begin{equation}
E_0(\beta, M) = E_0(M) + E'_0(M)/\beta.
\end{equation}
The right panel shows the continuum extrapolation
\begin{equation}
E_0(M) = E_0 + E'_0/M^2
\end{equation}
which is linear in $\epsilon^2$ since there are no discretization effects linear in $\epsilon$ in the
bosonic part of the action. We obtain the value $E_0 = 4.117(21)$ which 
should be compared with 4.117 resulting from the Hamiltonian approach.
We repeat the same procedure for the first excited state energy, $E_1$. Left panel of figure \ref{fig. e1}
shows the correlator $C_1(n)$ for different lattice spacings, which indeed is free from contributions coming
from higher excited states. From the slope of the correlator we read the value of $R(E_1-E_0)$, which 
we extrapolate to the infinite volume (see the right panel of figure \ref{fig. e1}). 
We obtain $E_1 - E_0 = 2.280(50)$, from which we calculate $E_1 = 6.397(71)$ to be compared with 6.388 
from the Hamiltonian approach.
%

\subsection{Full model}

Inspired by finite temperature lattice quantum field theory we perform simulations at 
a purely imaginary value of chemical potential at finite temperature \cite{gattringer} 
in order to be able to access sectors with a given fermionic occupation number in a lattice simulation.
The determinant of the Dirac operator at finite
temperature and finite chemical potential can be decomposed in a fugacity series
\begin{equation}
\det D(\mu) = \sum_q (e^{\mu \beta})^q D^{(q)}. \nonumber
\end{equation}
In the case of our non-relativistic quantum mechanics the sum runs over positive integer 
valued quark number $q \in [0,6]$.
The expansion coefficients $D^{(q)}$ are the canonical determinants and
may be obtained using Fourier transformation with respect to an
imaginary chemical potential,
\begin{equation}
D^{(q)} = \frac{1}{2\pi} \int_{-\pi}^{\pi} d\phi e^{-i q \phi}
\det D(\beta \mu = i \phi). \nonumber
\end{equation}
Similarly, for an expectation value of a bosonic observable $O$ we can write
\begin{equation}
\langle O \rangle^{(q)} = \frac{1}{2\pi} \int_{-\pi}^{\pi} d\phi e^{-i q \phi}
\langle O(\beta \mu = i \phi) \rangle. \nonumber
\end{equation}

\begin{figure}
\begin{center}
\includegraphics[width=0.485\textwidth]{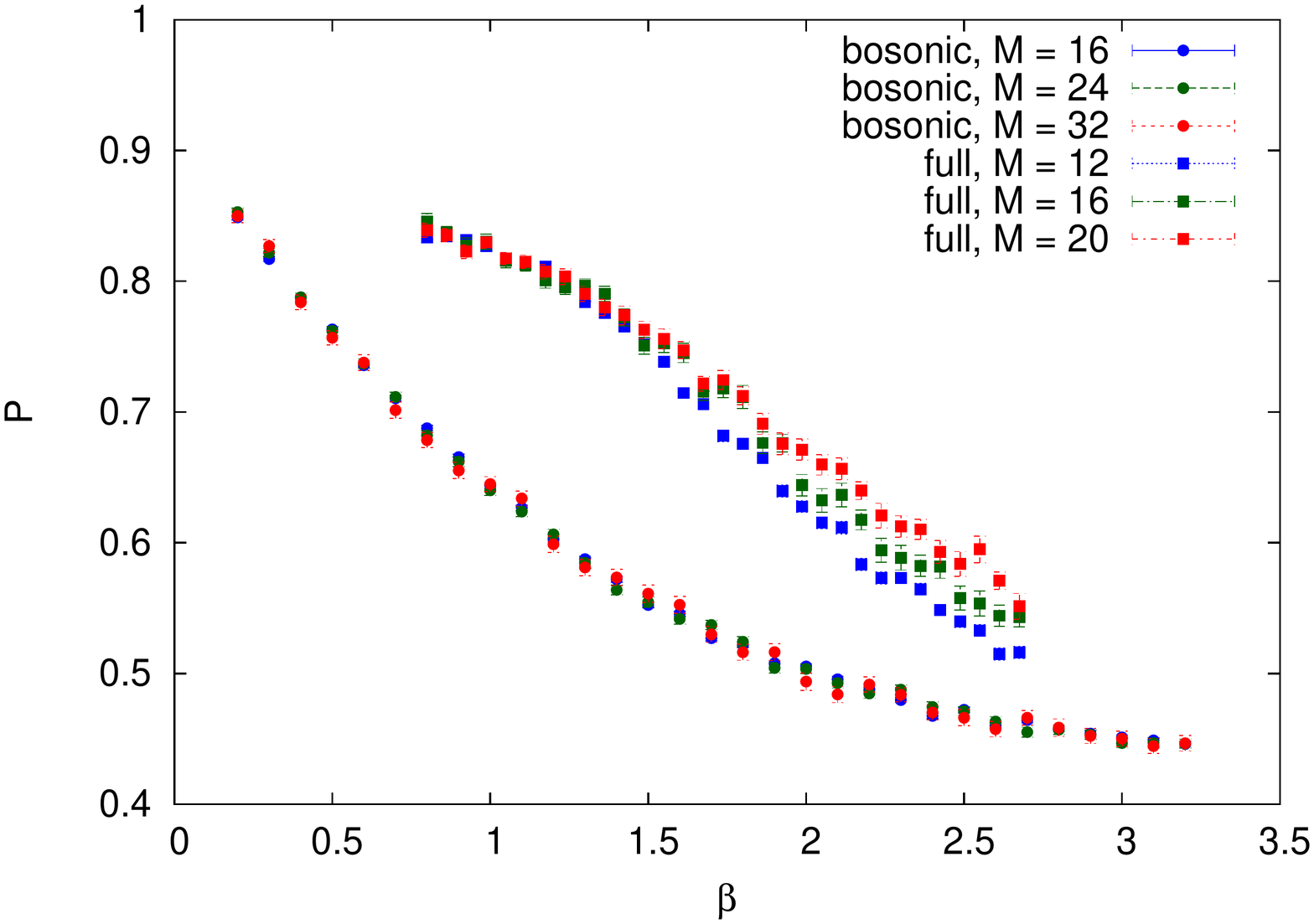}
\includegraphics[width=0.485\textwidth]{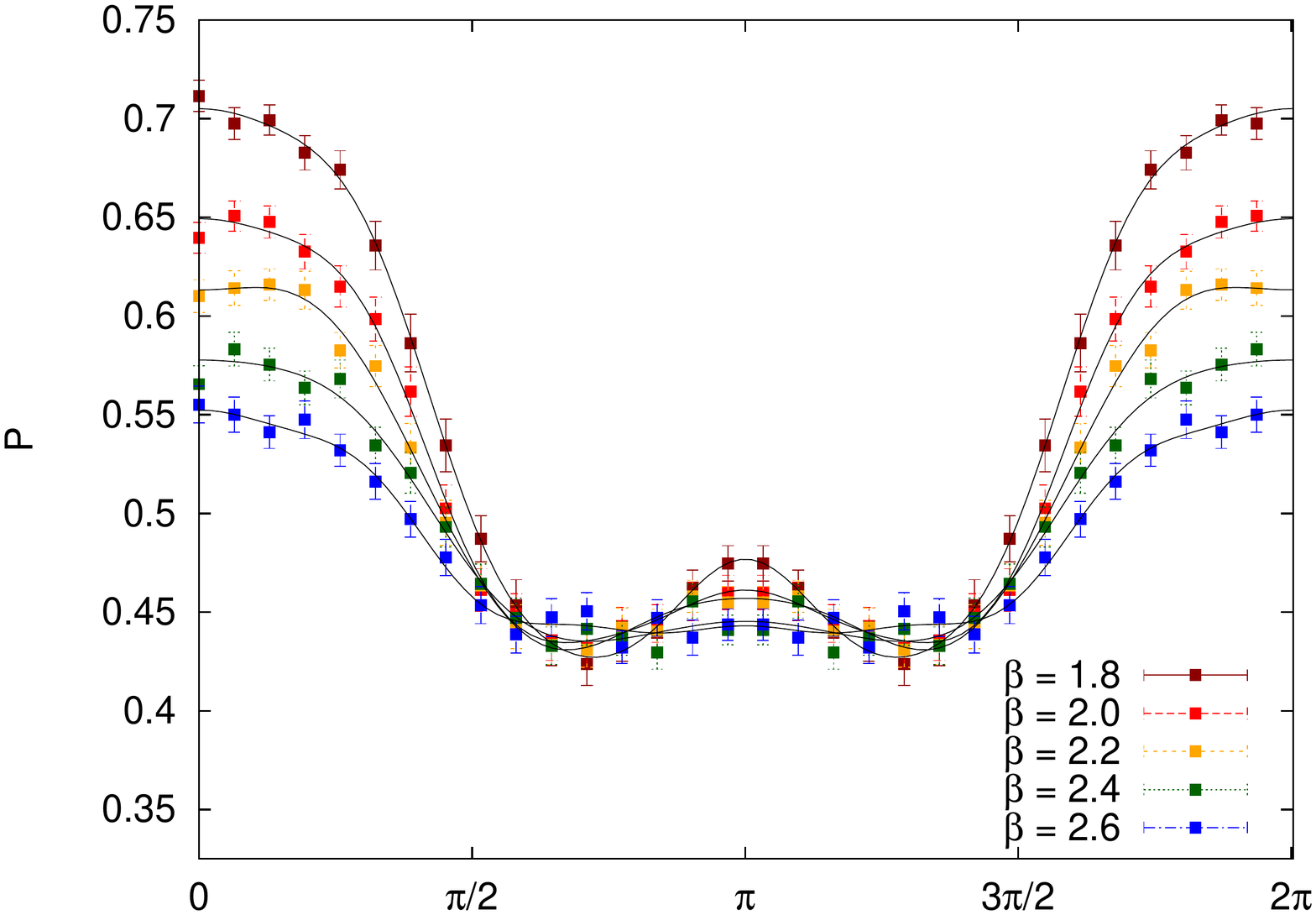}
\caption{$P$ as a function of temperature (left panel) and imagimary chemical potential (right panel)\label{fig. p}}
\end{center}
\end{figure}

\begin{figure}
\begin{center}
\includegraphics[width=0.5\textwidth]{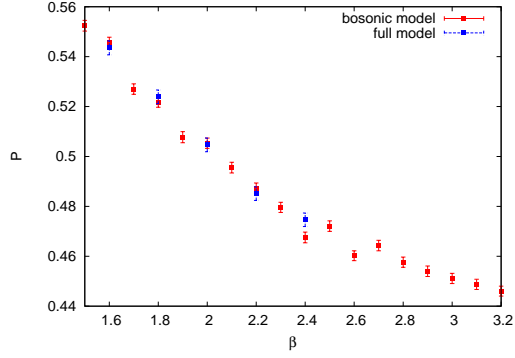}
\caption{Comparison of $P(\beta)$ calculated in the pure gauge model and the projection of $P(\beta \mu)$ calculated in the full model 
to the $n_F=0$ sector. \label{fig. p from mu}}
\end{center}
\end{figure}

In order to demonstrate the validity of such approach we calculated the expectation values of the Polyakov loop, $P$ as
a function of the temperature and chemical potential in both: the pure gauge model and the full, supersymmetric model. The left
panel of figure \ref{fig. p} shows the dependence of $P$ on $\beta$ for different lattice spacings. 
The lower sets of points correspond to the simulations in the bosonic sector only, whereas the upper set of points to the 
simulations in the full model. We notice that the bosonic results are close the their continuum value since there is 
no significant difference between data obtained on a 24 and 32 lattice. The lattices used to simulate the full model are smaller
and hence cut-off effects are visible for smaller temperatures (higher $\beta$). For a few selected values of $\beta$ we
performed simulations for different values of the imaginary chemical potential. The results are shown on the right panel
of figure \ref{fig. p}. The plot shows also a fit with the following functional ansatz
\begin{equation}
P(\beta \mu) = P^{(q=0)}(\beta) + \sum_{i=1}^6 \cos(\beta \mu) P^{(q=i)}(\beta) 
\end{equation}
from which we can extract the expectation value of the Polyakov loop in the sector without fermions, i.e. in the bosonic sector.
On figure \ref{fig. p from mu} we compare $P^{(q=0)}(\beta)$ with the direct determination of $P(\beta)$ already shown on 
figure \ref{fig. p} and notice a very good agreement.

\section{Conclusions}
\label{sec. conclusions}

We presented two independent non-perturbative approaches to supersymmetric Yang-Mills
quantum mechanics with 4 supercharges. Combined together they provide access to spectral 
and thermal properties of such models, which include the energies and quantum numbers of low-lying 
eigenstates and their wavefunctions. At this point one should mention a new promising 
formulation of the model discussed in this Letter in terms of fermion loops \cite{wenger}. 
Our further calculations include studies of the dependence on the imaginary chemical potential 
of other observables as well as models with higher SU(N) groups. 

\acknowledgments
P.K. thanks A. Ramos and A. Joseph for many discussions and M. Kore\'{n} for valuable discussions and for
writting an independent code. P.K. acknowledges support of NCN grant no. 2011/03/D/ST2/01932 and Z.A. of Foundation 
for Polish Science MPD Programme co-financed by the European Regional Development Fund, agreement no. MPD/2009/6.

\end{document}